\documentclass[5p,times]{elsarticle}

\begin{document}

\title{Phase stability of chromium based compensated ferrimagnets with inverse Heusler structure}

\author{Markus Meinert}
\ead{meinert@physik.uni-bielefeld.de}
\author{Manuel P. Geisler}
\ead{mgeisler@uni-bielefeld.de}
\address{Thin Films and Physics of Nanostructures, Department of Physics, Bielefeld University, D-33501 Bielefeld, Germany}

\journal{Journal of Magnetism and Magnetic Materials}

\begin{abstract}

Chromium based inverse Heusler compounds of the type Cr$_2$YZ (Y=Co, Fe; Z=Al, Ga, In, Si, Ge, Sn) have been proposed as fully compensated half-metallic ferrimagnets. Such materials are of large interest for spintronics because they combine small magnetic moment with high spin polarization over a wide temperature range. We assess their thermodynamic stability by their formation enthalpies obtained from density functional theory calculations. All compounds under investigation are unstable. Cr$_2$FeSi and Cr$_2$CoAl are stable with respect to the elemental constituents, but decompose into binary phases. Cr$_2$FeGe, Cr$_2$CoGa, Cr$_2$FeSn and Cr$_2$CoIn are found to be unstable with respect to their elemental constituents. We identify possible binary decompositions.

\end{abstract}

\maketitle

\section{Introduction}
Fully compensated half-metallic ferrimagnets (also known as half-metallic antiferromagnets) \cite{vanLeuken95} have promising properties for spintronics. Due to their internal spin compensation, the total moment is (nearly) zero at low temperature, yet the possibly strong local moments allow for high Curie temperatures. However, approaching the Curie temperature one expects to find a non-zero total magnetic moment \cite{Sasioglu09}. In conjunction with possible appearance of half-metallicity, such materials would be ideal electrode materials for spin transfer torque switching devices and other advanced applications \cite{Pickett01, Hu11}.

Among the inverse Heusler compounds (space group $F\bar{4}3m$, prototype Hg$_2$CuTi) \cite{Skaftouros13}, several compositions with 24 valence electrons have been identified which could have these unusual properties, in particular Cr$_2$CoGa and Cr$_2$FeGe \cite{Galanakis11}. In contrast to regular Heusler compounds, the inverse Heusler compounds do not possess inversion symmetry, such that the Cr atoms become nearest neighbors, which leads to an antiparallel coupling of their magnetic moments due to direct exchange interaction. These compounds prefer the inverted Heusler structure over the regular one and the local moments are predicted to be quite large. In consequence, high Curie temperatures of 1520\,K and 748\,K are expected, respectively \cite{Galanakis11}.

Experiments indicate that Cr$_2$CoGa crystallizes in a cubic structure, which might be the inverse Heusler structure \cite{Graf09}. However, magnetization measurements reveal that the Cr$_2$CoGa compound is susceptible to atomic disorder, which generates a non-zero moment \cite{Qian11,Hakimi13}. Further, the Curie temperature was observed to be only about 300\,K and the electrical resistivity was semimetallic \cite{Qian11}. Both experimental findings indicate that severe structural disorder must be present and that the actual crystal structure of Cr$_2$CoGa cannot simply be portrayed as the inverse Heusler structure. Recent experiments on Cr$_2$FeGe indicate that this compound crystallizes in a more complex tetragonal structure \cite{Hakimi13}.

These experimental results prompted us to perform a stability analysis of these compounds and their relatives of the type Cr$_2$YZ (Y=Co, Fe; Z=Al, Ga, In, Si, Ge, Sn) with 24 valence electrons. Our analysis is based on formation energies computed within density functional theory  (DFT) \cite{HK,KS}, similar to recent analyses for a large number of Heusler and inverse Heusler compounds given by Gille{\ss}en and Dronskowski \cite{Gillessen08, Gillessen09}.

\section{Computational approach}
The DFT calculations with the Perdew-Burke-Ernzerhof (PBE) \cite{PBE} exchange-correlation functional were performed with the \textsc{Elk} code \cite{elk}, an implementation of the full-potential linearized augmented plane-wave method (FLAPW). The muffin-tin sphere radii were set to 2.0\,bohr for all elements and the augmented plane-wave expansion was taken to $k_\mathrm{max} = 4.0$\,bohr$^{-1}$. $\mathbf{k}$-point meshes with $20 \times 20 \times 20$ points were used.

We obtained formation energies (zero-temperature formation enthalpies) with respect to the elemental solids from the total energies of the compounds and the total energies per atom of the constituent elements in their respective ground-state structures:
\begin{equation}\label{Eq1}
\Delta E_0\left( \mathrm{Cr_2YZ}  \right) = E_\mathrm{tot}^\mathrm{Cr_2YZ} - \left(  2 E_\mathrm{tot}^\mathrm{Cr}  + E_\mathrm{tot}^\mathrm{Y} + E_\mathrm{tot}^\mathrm{Z}  \right).
\end{equation}
The total energy differences were converged to within 10\,meV. Structural relaxations of the constituent elements were performed with the GPAW code \cite{gpaw,ase}. Decompositions into binary phases were studied with the help of the \textit{Materials Project} database \cite{MP1, MP2}.

\section{Results}


\begin{table*}[t]
\centering
\begin{tabular}{l r r r r r r r}
\hline\hline
			&	$a_0$ 		& $\Delta E_0$		&	$\Delta E_0$		&	$m_\mathrm{Y(A)}$ 		& 	$m_\mathrm{Cr(B)}$		&	$m_\mathrm{Cr(C)}$	&	$m_\mathrm{tot}$\\
			&	(\AA{}) 		& (eV/f.u.)			&	(kJ/mol)			&	($\mu_\mathrm{B}$) 		& 	($\mu_\mathrm{B}$)		&	($\mu_\mathrm{B}$)	&	($\mu_\mathrm{B}$)\\\hline
Cr$_2$FeSi	&	5.62			& $-$0.78			&	$-$75.6		&	      0.00				& 	0.00					&	    0.00			&	0.00\\
Cr$_2$FeGe	&	5.72			& 0.15			&	14.1			&	$-$0.22				& 	1.17					&	$-$0.92			&	0.04\\
Cr$_2$FeSn	&	6.07			& 1.13			&	108.8			&	$-$0.14				& 	2.11					&	$-$1.93			&	0.03\\
Cr$_2$CoAl	&	5.78			& $-$0.27			&	$-$26.0		&	      0.30				& 	1.36					&	$-$1.49			&	0.01\\
Cr$_2$CoGa	&	5.78			& 0.09			&	8.8			&	      0.40				& 	1.48					&	$-$1.64			&	0.09\\
Cr$_2$CoIn	&	6.06			& 1.06			&	101.8			&	      0.59				& 	2.06					&	$-$2.32			&	0.13\\
\hline\hline
\end{tabular}
\caption{\label{Tab1} Equilibrium lattice constants, formation energies in eV per formula unit and kJ per mole compound.  Site resolved local magnetic moments given in $\mu_\mathrm{B}$, where Y = Fe, Co and A, B, C, are described in the text.}
\end{table*}

\begin{figure}[t]
\includegraphics[width=8.6cm]{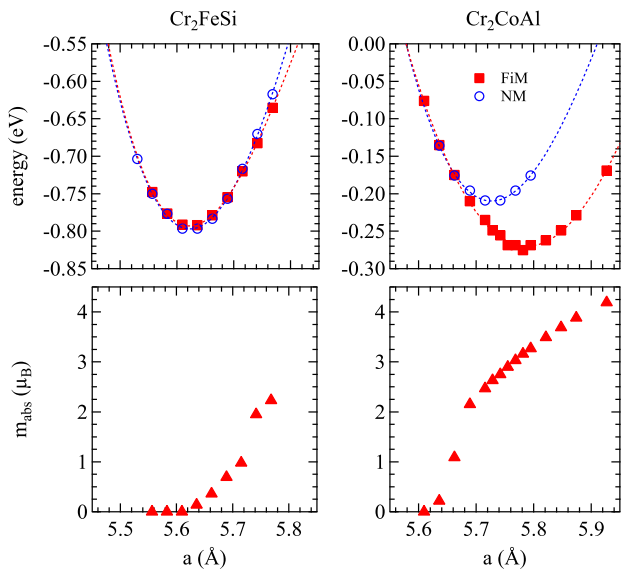}
\caption{\label{Fig1} Total energies for ferrimagnetic (FiM) and nonmagnetic (NM) configurations (top row) and absolute magnetic moments (bottom row) as functions of the lattice constant for Cr$_2$FeSi and Cr$_2$CoAl. Dotted lines represent third-degree polynomial fits.}
\end{figure}

The equilibrium lattice constants and formation energies and magnetic moments of the six compounds with 24 valence electrons are gathered in Table \ref{Tab1}. The total moments are nearly zero in all cases. The sites A, B, C, D, denote the internal coordinates $(0, 0, 0)$, $(\frac{1}{4}, \frac{1}{4}, \frac{1}{4})$, $(\frac{1}{2}, \frac{1}{2}, \frac{1}{2})$, and $(\frac{3}{4}, \frac{3}{4}, \frac{3}{4})$. Note that the values of the local magnetic moments depend strongly on the chosen muffin-tin sphere radii and may thus differ from values found in the literature. We find that only Cr$_2$FeSi and Cr$_2$CoAl are possibly stable in the inverse Heusler structure, since only these two compounds have negative formation energy. While Cr$_2$CoAl is ferrimagnetic, Cr$_2$FeSi is not. Cr$_2$FeSn and Cr$_2$CoIn are clearly thermodynamically unstable due to their large positive formation energy. Cr$_2$FeGe and Cr$_2$CoGa are critical cases, with the magnitude of the energy of formation close to zero. Thus, the quality of the prediction may suffer from the imperfect exchange-correlation functional and may be different for another functional. Gille{\ss}en and Dronskowski have estimated the accuracy of their calculations (which are similar to ours) by comparison with experimental data obtained on Ni$_3$Al, Ni$_3$Ga, and Ni$_3$Ge with AuCu$_3$ structure. They found deviations of $-4$ to $-12$\% ($-4$ to $-18$\,kJ/mol) from experiment, which are quite small. Experimentally, the trend is found that Cr$_2$CoGa is just stable with a disposedness to chemical disorder, while Cr$_2$FeGe is indeed not stable in the cubic phase and prefers a tetragonal one as described in the introduction.

\begin{table}[b]
\begin{tabular}{l r}
\hline\hline
reaction											& $\Delta E_0$ (eV/f.u.)\\\hline
3(2Cr + Fe + Si)		$\rightarrow$      2Cr$_3$Si + Fe$_3$Si		& -1.35\\
3(2Cr + Fe + Ge)		$\rightarrow$      2Cr$_3$Ge + Fe$_3$Ge	& -0.45\\
2Cr + Fe + Sn		$\rightarrow$	2Cr + Fe + Sn			& 0.00\\
2Cr + Co + Al		$\rightarrow$	2Cr + CoAl				& -1.20\\
2Cr + Co + Ga		$\rightarrow$      2Cr + CoGa			& -0.56\\
3(2Cr + Co + In)		$\rightarrow$	In$_3$Co + 6Cr + 2Co		& -0.05	      \\
\hline\hline
\end{tabular}
\caption{\label{Tab2} Most favorable reactions forming binary phases and their reaction energies according to the \textit{Materials Project} database. There are no stable binary phases in the Cr-Fe-Sn system.}
\end{table}


For comparison we have calculated the formation energies of the Co$_2$FeSi, Co$_2$FeGe, Co$_2$FeSn series of full Heusler compounds. We find $-1.38$, $-0.59$, and $+0.05$\,eV/f.u., respectively. Indeed, Co$_2$FeSn is known to be thermodynamically unstable and to disintegrate into binary phases, but it can be synthesized in the Heusler phase in a non-equilibrium process \cite{Tanaka12}. Its binary decomposition into CoFe + CoSn has a formation energy of $-0.32$\,eV/f.u. In contrast, Co$_2$CrSi has a formation energy of $-0.85$\,eV/f.u., yet it is found to be unstable and to form disordered and binary phases \cite{Aftab11}. The most favorable binary decomposition is Co$_2$Si + Cr, with a formation energy of -1.35\,eV/f.u.

Searching for possible binary decompositions of the Cr based compounds, we find that also Cr$_2$CoAl and Cr$_2$FeSi are unstable. Cr$_2$CoAl would most likely decompose into 2Cr + CoAl ($\Delta E_0 = -1.20$\,eV/f.u.), whereas 3Cr$_2$FeSi would decompose into 2Cr$_3$Si + Fe$_3$Si ($\Delta E_0 = -1.35$\,eV/f.u.). The most favorable binary reactions for the six compounds are given in Table \ref{Tab2}.

For completeness, we study the magnetism as a function of the lattice constant of Cr$_2$FeSi and Cr$_2$CoAl in Fig. \ref{Fig1}, where the absolute moment denotes the sum of the absolute values of muffin-tin moments. The energy zero is chosen as the sum of the elemental total energies, i.e., the second term on the right-hand side of Eq. \ref{Eq1}. Cr$_2$FeSi is nonmagnetic at its equilibrium lattice constant, but turns into a weak ferrimagnet at slightly increased volume. Cr$_2$CoAl has a ferrimagnetic ground-state in the inverse Heusler structure. Due to the similarly large magnetic moments as in Cr$_2$CoGa, a high Curie temperature may be expected for this compound \cite{Galanakis11}.

\section{Conclusions}
In summary, we have studied the energy of formation for six chromium based ternary materials with inverse Heusler structure, which potentially could form compensated ferrimagnets. We have shown that all of these compounds are thermodynamically unstable, having positive formation energy or because of energetically more favorable binary decompositions. However, the inverse Heusler structure may be metastable at finite temperature, eventually opening a path to the synthesis of the compounds we have studied.

\section*{Acknowledgements}
We thank the developers of the \textsc{Elk} and GPAW codes and of the \textit{Materials Project} for their efforts. We further thank G\"unter Reiss for his support. 
Financial support by the Deutsche Forschungsgemeinschaft (DFG) is gratefully acknowledged.

\end{document}